\documentclass[preprint,12pt]{elsarticle}

\usepackage{amssymb}
\journal{Chaos}

\begin{document}

\begin{frontmatter}

%% Title, authors and addresses

%% use the tnoteref command within \title for footnotes;
%% use the tnotetext command for theassociated footnote;
%% use the fnref command within \author or \address for footnotes;
%% use the fntext command for theassociated footnote;
%% use the corref command within \author for corresponding author footnotes;
%% use the cortext command for theassociated footnote;
%% use the ead command for the email address,
%% and the form \ead[url] for the home page:
%% \title{Title\tnoteref{label1}}
%% \tnotetext[label1]{}
%% \author{Name\corref{cor1}\fnref{label2}}
%% \ead{email address}
%% \ead[url]{home page}
%% \fntext[label2]{}
%% \cortext[cor1]{}
%% \affiliation{organization={},
%%             addressline={},
%%             city={},
%%             postcode={},
%%             state={},
%%             country={}}
%% \fntext[label3]{}

\title{ Quantitative Analysis of Social Influence \& Digital Piracy Contagion with Differential Equations on Networks}

%% use optional labels to link authors explicitly to addresses:
%% \author[label1,label2]{}
%% \affiliation[label1]{organization={},
%%             addressline={},
%%             city={},
%%             postcode={},
%%             state={},
%%             country={}}
%%
%% \affiliation[label2]{organization={},
%%             addressline={},
%%             city={},
%%             postcode={},
%%             state={},
%%             country={}}

\author[inst1]{Dibyajyoti Mallick}

\affiliation[inst1]{organization={Department of Physics},%Department and Organization
            addressline={National Institute of Technology Durgapur}, 
            city={Durgapur},
            postcode={713209}, 
            state={West Bengal},
            country={India}}

\author[inst2]{Kumar Gaurav}
\author[inst3]{Saumik Bhattacharya}
 \author[inst1]{Sayantari Ghosh}
\affiliation[inst2]{organization={Department of Electronics Engineering},%Department and Organization
            addressline={Harcourt Butler Technical University}, 
            city={Kanpur},
            postcode={208002}, 
            state={Uttar pradesh},
            country={India}}
            \affiliation[inst3]{organization={Department of Electronics and Electrical Communication Engg.},%Department and Organization
            addressline={Indian Institute of Technology Kharagpur}, 
            city={Kharagpur},
            postcode={721302}, 
            state={West Bengal},
            country={India}}

\begin{abstract}
%% Text of abstract
Though the studies of social contagions are regularly borrowing network models to study the propagation of social influences and opinions to include social heterogeneity. Such studies provide valuable insights regarding these, but the social network structures cannot be well explored in their study. In this research, we methodically study the trends in online piracy with a continuous ODE approach and differential equations on graphs, to have a clear comparative view. %This study includes sharing files through peer-to-peer networks, downloading or streaming content from illegal websites, or sharing access to paid subscription services, traditional industry supply chains, software, Music, print media, etc, entertainment platforms are some of the most vulnerable content pirated online and its rapid growth nowadays signifies it as a social epidemic.
We first formulate a compartmental model to mathematically study bifurcations and thresholds, and later move on with a network-based analysis to illustrate the proliferation of online piracy dynamic with an epidemiological approach over a social network. We figure out a solution for this online piracy problem by developing awareness among individuals by introducing media campaigns which could be a useful factor for the eradication and control of online piracy. Next, using degree-block approximation, network analysis has been performed to investigate the phenomena from a heterogeneous approach and to derive the threshold condition for the persistence of piracy in the population in a steady state. Based on the behavioral responses of individuals in a society due to the effect of media, we examine the system through the aid of realistic parameter selection to better understand the complexity of the dynamics and propose control strategies.
\end{abstract}

%%Graphical abstract
%\begin{graphicalabstract}
%\includegraphics{grabs}
%\end{graphicalabstract}

%%Research highlights
%\begin{highlights}
%\item Research highlight 1
%\item Research highlight 2
%\end{highlights}

\begin{keyword}
%% keywords here, in the form: keyword \sep keyword
Online social epidemics \sep Online Piracy \sep Information diffusion and propagation \sep Mean-field analysis and bifurcations \sep Graph-theoretical treatment \sep Epidemic eradication strategies.
%% PACS codes here, in the form: \PACS code \sep code
\PACS 0000 \sep 1111
%% MSC codes here, in the form: \MSC code \sep code
%% or \MSC[2008] code \sep code (2000 is the default)
\MSC 0000 \sep 1111
\end{keyword}

\end{frontmatter}

%% \linenumbers

%% main text
\section{Introduction} \label{sec:introduction}
Internet-based spreading mechanisms such as tweeting and sharing online
content, like, rumors, gossip, hoaxes, campaigns, etc. have drawn the attention of research communities in the last decade, due to their striking similarities with the viral spread of disease, causing epidemic \cite{zhu2017rumor, gonccalves2019optimal}. Starting from the very initial SIR models \cite{kermack1932contributions,kabir2019analysis}, mathematical epidemiology has grown over time, with contributions from various disciplines, like, applied mathematics, computer science, social science, and graph theory \cite{boccaletti2020modeling,burgess2018influence,rodrigues2016can}. The assumptions associated with the classical compartmental models, like homogeneous mixing, have been identified as questionable, especially, in the case of social contagions, due to the exceptionally organized nature of online social networks. Although, these models are efficient in providing answers to the questions like how many people are affected by a given social contagion at any given time; how many will remain infected in a steady state; what is the nature of bifurcation, etc., they are silent on the questions, like, who are expected to get affected at a particular time; who are the critical persons for controlling the contagion flow. In short, all population-level statistics are provided by these models, but information related to a particular individual is missing \cite{bansal2007individual,hayes2010psoriasis}. Other than some recent exceptions \cite{gaurav2017equilibria, bhattacharya2019viral,apolloni2014metapopulation}, particular case studies on the comparison of homogeneous and heterogeneous approaches of social epidemic models is a less explored area at this point. \\
In many studies, it is found that young people are more likely to engage in online piracy as an acquired habit due to a lack of awareness of copyright laws and easy access to the internet. However, the perception that piracy is a victimless crime %even in the presence of their sense of moral obligation 
\cite{sinha2008preventing,nandedkar2012won,yoon2011theory,ramayah2009testing, lee2011understanding,khang2012exploring,larsson2013online} is the most powerful reason behind this, and the reason for this perception is that usually they are introduced to piracy through their peers. %While this negative peer influence brings them into the arena of piracy, there is also the possibility of the existence of positive peer influence.  Friends and well-wishers sometimes highlight the fact to youngsters that online piracy is illegal and can have serious consequences. It is also important to recognize that piracy can harm the creative industries.//
To address these moral perspectives, several approaches, like, education, law enforcement, media campaign, and awareness campaigns can be effective, by informing people about the risks and consequences of piracy. Additionally, making content more affordable and accessible through legal channels (like, OTT platforms, etc.) may reduce the temptation to pirate and also be used as a control strategy. \\
%Another aspect of media campaigns could relate to associated consequences. %Several researchers  \cite{nandedkar2012won,yoon2011theory,chen2008intention} have shown that the implementation of punishment can also be one of the prime factors that reduce the habit of piracy. In a recent survey \cite{irdeto2017nearly} with more than 25000 participants, 48\% of people informed that information about damages incurred by piracy and probable punishments will affect their decision. In \cite{wang2013drives}, authors mentioned that momentary enjoyment and a sense of benefit from piracy are always gets dominated by ethical efficacy and fear. Several independent research findings pointed out some leading reasons for individuals, like, the information-dependent moral psychology that controls the habit of piracy \cite{moores2006ethical,coyle2009buy}. \\
%So, both these aspects, legal as well as ethical concerns related to piracy can be evoked using external influences of media campaigns as well as using word-of-mouth awareness \cite{danaher2014gone,cronan2008factors}.\\
In recent work, \cite{gaurav2022purchase}, a model of peer influence for piracy has been studied, where they have studied word-of-mouth awareness spreading, against piracy, as a possible way to curb it. However, the holistic social enlightenment we mentioned here cannot be modeled using the conventional one-to-one awareness model. Rather, it requires a model with more complexity and nonlinearity that includes society-wide awareness campaigns by government agencies \cite{paul2021covid,misra2011modeling,samanta2013effect}. Moreover, assuming a simple ODE model with homogeneous mixing is only effective to observe a few critical dynamics of complex systems such as steady states, effects of the various modeling parameters, sensitivity, etc. It does not cogitate a realistic scenario where a person can interact with a limited specific number of other people, rather than randomly. So, to account for the heterogeneity and individual-level information, we plan to study this spreading of epidemic-like piracy on networks \cite{viguerie2021simulating,kabir2020impact,espinel2012combating,sharma2014modeling}. We will modify the differential equations for networks, using graph theoretical tools, %considering each person is a node and the relation between them is a link in the network. Though, both deterministic and network modelings give crucial insights into the habit of using pirated products, the relation between these two methods needs to be explored. In this paper, we try to relate these two methods to understand the effectiveness of both methods in predictive modeling. 
assuming that each individual is a node and that the connections between them form links in the network. Deterministic and network models both shed light on the practice of using counterfeit goods, but the relationship between the two approaches needs to be investigated. In this study, we attempt to relate these two approaches in order to comprehend how effective each approach is for predictive modeling.
The respective compartmental model with media campaign is proposed in Section II and analyzed the bifurcations and thresholds mathematically through a homogeneous approach \cite{li2009stability,ccakan2020dynamic,lemaitre2021scenario,sharma2015backward}, afterward move on with a network-based analysis to illustrate the proliferation of online piracy dynamic over a social network through heterogeneous approach and the numerical simulations on the model as well as a real jazz network is being discussed in Sections III, IV, and V respectively. The results are concisely discussed and summarized in Section VI.
 
\section{Proposed Model with Mass Media Awareness}
\label{model_5}
\begin{figure}
\begin{center}
%\begin{minipage}[c]{1\columnwidth}
\includegraphics[width=0.7\linewidth]{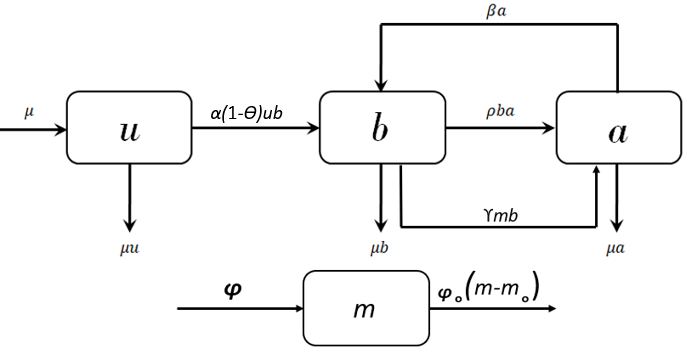}
%\end{minipage}
\caption{Block diagram of the proposed model for the propagation of habit of online piracy in the presence of mass media awareness.}
\label{block_3M}
\end{center}
\end{figure}

To create a model of the piracy dynamics, we considered our compartmental model, with a total population of $T$, and categorized each individual into three compartments: Unaware (U), Bootleggers (B), and Aware (A). The unaware class, denoted by U, is yet to be habituated to online piracy; these are susceptible people, who may get influenced by bootleggers with rate $\alpha$ and join them in such activities. The bootlegger group of people B consists of individuals who have these online piracy habits and spread these habits through their social contacts. Finally, we have the aware class A, who used to be bootleggers but they left the habit of piracy at present and becoming aware of the negative repercussions of it. They spread awareness against piracy habits among bootleggers and influence them to join class A with the effective rate $\rho$. Meanwhile, a fraction of class $A$ who can not resist themselves to use certain  
pirated content can again join B class with a relapse rate of $\beta$.  For demography, birth and death rates $\mu$ have been considered here, which effectively maintains a fixed population size. 
Considering the total population $T= 1$ (normalized form),  
 \begin{equation}
   u+b+a = 1   
 \end{equation}
Now, let us consider the external efforts, like global initiatives by Governments or international law enforcement organizations that are made to bring down the severity of online piracy. Instead of distinguishing different parameters like media campaigns, law enforcement, punishments, advertisements, social drives to boost ethical and moral values, etc., we combine them in a single time-varying parameter, and for the sake of simplicity, we will define it as `\textit{an effect of media}'. The Cumulative density of awareness programs driven by the media is denoted by $m$. The rate of change of media is given by the differential equation.
$
m'=\phi b- \phi_{0}(m-m_{0})\cdot  
$
As clear from this equation, the extent of media program implementation is assumed to be increasing linearly with the proportion of bootleggers in the population at the rate $\phi$. The depletion rate of these programs due to the ineffectiveness of social barriers in the population is $\phi_{0}$. The awareness level of the society before the implementation of any awareness program is denoted by $m_{0}$. It shall be noted that $m$ is always higher than $m_0$. When $m =m_0$, then $m' = \phi b$.
\newline
The effect of media awareness on bootleggers has been incorporated in the model by introducing a transition from bootlegger to aware at the rate $\gamma m$ where $\gamma$ is the success rate at which a person moves from class $B$ to class $A$. In the presence of an external awareness program, the conversion rate from unaware to bootleggers also decreases. This reduction has been incorporated by multiplying $\alpha $ by a factor $(1-\Theta)$ where $\Theta$ is $\frac{m}{c+m}$. The positive constant $c$ limits the effect of awareness programs on unawares and is known as the half-saturation point for Holling type-II functional response \cite{greenhalgh2015awareness}.
A Schematic diagram representing all the transitions between different classes has been shown in Fig. \ref{block_3M}. The coupled differential equations for the model are as follows:
\begin{eqnarray}
u'&=&\mu -\alpha  (1-\Theta) u b -\mu u ,\nonumber \\
b'&=&\alpha  (1-\Theta) u b-\rho b a+\beta a -\gamma m b -\mu b ,\nonumber \\
a'&=&\rho b a-\beta a +\gamma m b - \mu a , \label{diff_eq:uba_with_media_model_3} \\
m'&=&\phi b- \phi_{0}(m-m_{0})  \cdot \nonumber
\end{eqnarray}
\section{Homogeneous Analysis}
\subsection{Equilibrium Analysis}
After attaining the steady state, the rate of change of all the subpopulations that is u, b, and a must be zero. Solving the eq. set \ref{diff_eq:uba_with_media_model_3}, to zero, we get        
\begin{equation}
m^\star=m_{0}+zb^\star ; \; 
u^\star={\scriptstyle{ \frac{(c+m_{0}+zb^\star)\mu}{(c+m_{0}+zb^\star)\mu+\alpha  c b^\star}}} ; \; 
a^\star={\scriptstyle{\frac{(m_{0}+zb^\star)\gamma b^\star}{\beta+\mu-\rho b^\star}}}
\label{u_and_a_in_b}
\end{equation}
where $z=\frac{\phi}{\phi_{0}} \cdot$
Substituting the values of $u^\star$ and $a^\star$ in $(u^\star+b^\star+a^\star=1)$,
we get a cubic equation in $b^\star$ i.e.,\\ \\
 $p (b^{\star})^{3}+q (b^{\star})^{2}+r b^\star=0$ 
where,
 \begin{eqnarray}
p&=&(\mu z+\alpha  c)(\gamma z-\rho), \nonumber \\
q&=&(\mu z+\alpha  c)(\beta +\mu+m_{0} \gamma)+\mu(c+m_{0})(\gamma z-\rho)%\nonumber \\&\phantom{{}=1}&
+\alpha  c \rho, \nonumber  \\
r&=&\mu(c+m_{0})(\beta +\mu+m_{0} \gamma)-\alpha  c (\beta+\mu) \cdot \label{coeff_pqr_with_media_model_3}
 \end{eqnarray}
 From the equation, it can be observed that $b^\star=0$ is always a solution. The respective values of $u^\star$ and $a^\star$ are 1 and 0. Hence, $E_{0}(1, 0, 0)$ is always a steady-state solution of the system. To investigate other roots, we consider $b^\star \neq 0$ and are left with quadratic equation $p (b^{\star})^{2}+q b^{\star}+r = 0$, where on the basis of sign of the coefficient $p$ there are two different situations: \\
 \textbf{Case 1}: $p>0$ implies $\gamma z>\rho$, which in turn also implies $q>0.$ There will be one (or no) positive solution depending on whether $r$ is negative (or positive). Hence, the required condition to have a positive solution of the quadratic equation is
 \begin{equation}
\alpha  c (\beta+\mu)> \mu(c+m_{0})(m_{0} \gamma+\beta +\mu)\cdot \nonumber
\end{equation}
It gives the expression of reproduction number $\mathcal{R}_{m}.$ A single endemic steady state exists only when 
\begin{equation}
\mathcal{R}_{m}=\frac{\alpha  c (\beta+\mu)}{\mu(c+m_{0})(\beta +\mu+m_{0} \gamma) }>1 \cdot
\end{equation}
otherwise, only a piracy free steady state $E_{0}$ exists. \\
\textbf{Case 2}: When $p<0$ i.e, $\gamma z<\rho.$ Depending on the sign of $r$ this case can be further divided into two sub-cases. \\
\textbf{Sub-case 2.1}: When $r>0$ ($\mathcal{R}_{m}<1$), similar to Case 1, one of the roots is positive, and the other is negative.  %We need to figure out whether the positive root gives a feasible steady state point or not.
\\
\textbf{Sub-case 2.2}: When $r<0$ ($\mathcal{R}_{m}>1$),
$q$ will surely be positive. It can be observed after a bit of rearrangement in the expression of $q$ and $r$ given in eq. set  \ref{coeff_pqr_with_media_model_3}.
\begin{eqnarray}
q&=& (\mu z+\alpha  c)(\beta +\mu+m_{0} \gamma)+\mu(c+m_{0})\gamma
z\nonumber \\ &\phantom{}& - \rho (\mu c+\mu m_{0}-\alpha  c), \nonumber \\
r&=& \mu(c+m_{0})(m_{0} \gamma)+(\beta+\mu)(\mu c+\mu m_{0}-\alpha  c)\cdot \nonumber
\end{eqnarray}
The term is responsible for a negative sign of $r$ i.e., $(\mu c+\mu m_{0}-\alpha c)$ is appearing in the expression of $q$ with a negative sign resulting in positive $q$. With $p<0$, $q>0$, and $r<0$ it is ensured that both roots of the quadratic equation will be positive. %We need to figure out whether both these steady states are feasible or not.
\\
Quadratic equation in $b^{\star}$ tells that there will be one or two positive solutions when ($\mathcal{R}_{m}<1$) or ($\mathcal{R}_{m}>1$), but it does not guarantee that all these solutions will be physical (i.e., $0\leq b^\star \leq 1$). To investigate it further, along with the quadratic equation in $b^\star$,  we have also analyzed a quadratic equation in $a^\star$.  At a steady state, the rate of change of all three classes will be zero. Equating the third equation of the coupled eq.  \ref{diff_eq:uba_with_media_model_3} to zero, we get
\begin{equation}
\rho b a-\beta a +\gamma m b - \mu a =0 \cdot \nonumber
\end{equation}
At steady state, value of $b$, $a$, and $m$ will be $b^\star$, $a^\star$, and $m^\star$ respectively. Replacing $b^\star$ by  $(1-u^\star-a^\star)$ and substituting $m^\star$ and $a^\star$ from eq. \ref{u_and_a_in_b}, we get a quadratic equation $p_{a}(a^\star)^{2}+q_{a}a^\star+r_{a}=0$ where
\begin{eqnarray}
p_{a}&=& (\gamma z-\rho),\nonumber\\
q_{a}&=& (\rho-2z\gamma)(1-u^\star)-\gamma m_{0} -(\mu+\beta),\\
r_{a}&=& \gamma m_{0}(1-u^\star)+\gamma z-\gamma z u^\star (2-u^\star)\cdot \nonumber
\label{coeff_paqara}
 \end{eqnarray}
Based on the above discussion, we conclude that there exists only one physical endemic steady state beyond  $\mathcal{R}_{m}>1$.\\
For $\mathcal{R}_m<1$, examining the roots of quadratic equations in $b^\star$ and $a^\star$, it can be observed that one of the roots is positive and another is negative in both cases. For any physical solution $u^\star$, $b^\star$, and $a^\star$ must be in the range [0,1] individually. For $0\leq u^\star <1$, quadratic equation in $b^\star$ and $a^\star$ gives two roots ($b^{\star}_1, b^{\star}_2$) and ($a^{\star}_1, a^{\star}_2$) respectively which finally form steady state triplets $(u^{\star}_1, b^{\star}_1, a^{\star}_1)$ and $(u^{\star}_2, b^{\star}_2, a^{\star}_2)$. For $\mathcal{R}_m<1$, if $b^{\star}_1$ is positive then $b^{\star}_2$ is negative or vice versa. As discussed earlier, the statement also holds for $a^{\star}_1$ and $a^{\star}_2$ in this range. To check the feasibility of the triplets, let us assume that $(u^{\star}_1, b^{\star}_1, a^{\star}_1)$ is physical, i.e., $u^{\star}_1$, $b^{\star}_1$, and $a^{\star}_1$ are in the range $0$ to $1$ individually, and $u^{\star}_1+b^{\star}_1+a^{\star}_1=1$. However, if our assumption is true, then it implies that $b^{\star}_2$ and $a^{\star}_2$ will surely be negative. But, for  $0\leq u^{\star}_2 < 1$, along with negative values of $b^{\star}_2$ and $a^{\star}_2$, the required condition of $u^{\star}_2+b^{\star}_2+a^{\star}_2=1$ will never hold. Hence, our initial assumption that both $b^{\star}_1$ and $a^{\star}_1$ are positive, is not right. Actually, for positive  $b^{\star}_1$, $a^{\star}_1$ will be negative and for negative $b^{\star}_1$, $a^{\star}_1$ will be positive. Hence, we conclude that for $\mathcal{R}_m<1$, neither of the endemic states is physical and only piracy free steady-state prevails.
\newline
We conclude that the scenario is the same whether $\gamma z <\rho$ or $\gamma z > \rho$. For $\mathcal{R}_{m}<1$ there exists only a piracy-free steady state and for $\mathcal{R}_{m}>1$ there exists a unique endemic steady state.\\
\textbf{Sensitivity Analysis:}
The value of $\mathcal{R}_m$ depends on $\alpha $, $c$, $\beta$, $\mu$ $m_0$, and $\gamma$. For any parameter $x$ the sensitivity of $\mathcal{R}_m$ is defined as
\begin{equation}
\zeta^{\mathcal{R}_m}_{x}=\frac{x}{\mathcal{R}_m}\cdot \frac{\partial \mathcal{R}_m}{\partial x}
\end{equation}
Using this definition, sensitivities of $\mathcal{R}_m$ for all six parameters are as follow:
\begin{eqnarray}
\zeta^{\mathcal{R}_m}_{\alpha }&=&1,\;\;\;
\zeta^{\mathcal{R}_m}_{c}=\frac{m_0}{c+m_0}, \nonumber\\
\zeta^{\mathcal{R}_m}_{\beta}&=&\frac{\beta}{\beta+\mu}\cdot \frac{m_0 \gamma}{\beta+\mu+m_0 \gamma}, \nonumber\\
\zeta^{\mathcal{R}_m}_{m_0}&=&-(\frac{m_0 \gamma}{\beta+\mu+m_0 \gamma}+ \frac{m_0}{c+m_0}), \nonumber \\
\zeta^{\mathcal{R}_m}_{\mu}&=&-(\frac{\beta}{\beta+\mu}+\frac{\mu}{\beta+\mu+m_o \gamma}), \nonumber \\
\zeta^{\mathcal{R}_m}_{\gamma}&=&-\frac{m_0 \gamma}{\beta+\mu+m_0 \gamma}\cdot \nonumber
\end{eqnarray}
\subsection{Bifurcation and Physical Interpretation}
\label{stability_3A}
\begin{figure}
\begin{center}
%\begin{minipage}[c]{1\columnwidth}
\includegraphics[width=0.95\linewidth]{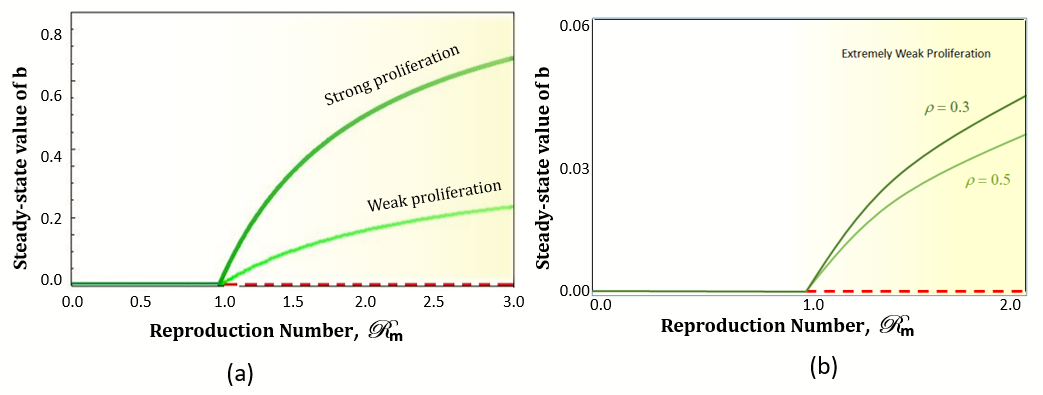}
%\end{minipage}
\caption{Variation of reproduction number $\mathcal{R}_{m}$ with the steady-state fraction of $b$ (a) Media effect: green line represents no media while the light green line shows the presence of media. (b) For two different values of $\rho$ i.e., 0.3 and 0.5. 
Other parameter values are $\mu=0.05$, $\beta = 0.2$, $\gamma = 0.08$, $\phi = 0.05$, $\phi_{0}= 0.01$, $c=5$ and $m_{0}=4$.
In these figures, green lines indicate stable solutions and red lines indicate unstable solutions.}
\label{bifurcation_3A}
\end{center}
\end{figure}

The bifurcation diagram for the system is shown in Fig.\ref{bifurcation_3A}. A forward transcritical bifurcation is observed at $\mathcal{R}_{m}=1$, where a new endemic steady state appears and the endemic free equilibrium loses its stability. For comparison, we also analyze the bifurcation for $m=0$ and $\phi=0$ as shown in fig \ref{bifurcation_3A}(a). In terms of physical interpretation, the most important difference that we observe here is the drastic decrease in the bootlegger population at the steady state. In the presence of mass media campaigns, the value of $b^{\star}$ becomes nearly 15\% for the considered parameter values which are very small compared to the model without media. With media, the fraction of people interested in piracy, even for higher values of reproduction number, remains substantially less, which can reduce the adverse effects of the piracy epidemic successfully. We denote this mass-level control strategy as \textit{weak proliferation}, which comes out to be an excellent and realistic way to handle the piracy epidemic. The bifurcation diagram for two different values of $\rho$ has been shown in Fig. \ref{bifurcation_3A}(b) and it can be observed that the steady-state fraction of bootleggers decreases further with an increase in awareness parameter $\rho$. This establishes that for a habit like online piracy, which has a very high degree of induction due to straightaway economic benefits to the users this mass-media-driven law enforcement seems to be a very efficient strategy of control. 
\subsection{Effect of Mass Media}
As discussed in the previous section, system behavior depends on the value of $\mathcal{R}_{m}$, which in turn depends on other parameters of the model. To understand the effect of media on the process, we have investigated the individual contribution of parameters $c$ and $m_{0}$ which control the effect of the media campaign.
\subsubsection{Effect of $c$}

\begin{figure}
    \centering
    \includegraphics[width=1\linewidth]{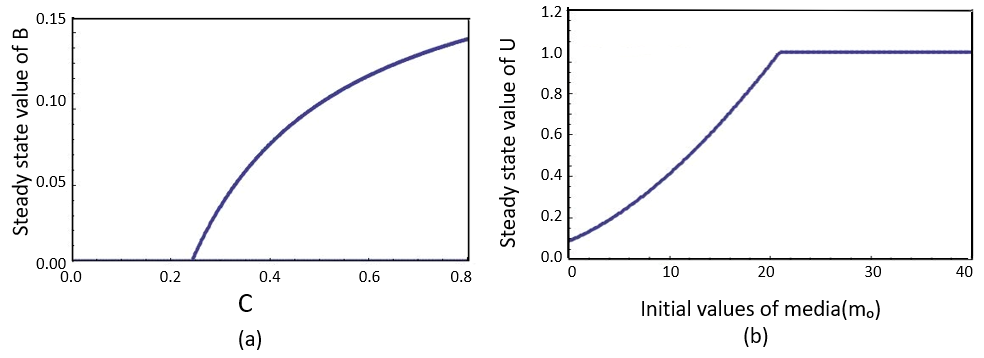}
  \caption{ Variation in steady state fraction of (a) bootleggers with constant $c$ which in turn changes the value of $\Theta$ and decides the rate of conversion from class $U$ to $B$ and (b) unaware class with the different initial level of intrinsic social awareness $m_0$. Parameters having same value in both the cases are $\mu=0.05$, $\alpha =2$, $\beta=0.2$, $\rho=0.3$, $\gamma=0.08$, $\phi=0.5$, and $\phi_{0}=0.1$. In (a) $m_0=4$ and in (b) $c=5$.  Threshold $c_{th}$ in (a) is 0.24178 and $m_{0_{th}}$ in (b) is 20.955.}
    \label{fig:bif steady}
\end{figure}

The parameter $c$ controls the value of $\Theta$ which in turn modifies the rate of conversion from class $U$ to class $B$ in the presence of mass media awareness. For piracy habit to extinct from the population, the required condition is $\mathcal{R}_{m}<1$ i.e.,
$
\mathcal{R}_{m}=\frac{\alpha  c (\beta+\mu)}{\mu(c+m_{0})(\beta +\mu+m_{0} \gamma) }<1 
$.
Rearranging the above equation, we get the threshold condition on the  value of $c$ as
$
c<\frac{m_{o}\mu}{(\alpha -\mu)-(\frac{\alpha  m_{0}\gamma}{\beta+\mu+m_{0} \gamma})}$.
 The right-hand side of the inequality is denoted by $c_{th}$.
 As shown in Fig. \ref{fig:bif steady}(a), for $c<c_{th}$ value of $b$ is 0, which means that the complete population is in an endemic free state and the problem of piracy does not exist. For $c$ greater than $c_{th}$, a non-zero value of $b$ can be observed, signifying the presence of people indulging in the habit of piracy.
\subsubsection{Effect of $m_{0}$}
Analyzing the effect of intrinsic level of social awareness $m_{0}$, we get the following quadratic expression from the threshold condition $\mathcal{R}_{m}<1$
\begin{equation}
\mu \gamma m_{0}^2+\mu(\beta+\mu+c\gamma)m_{0}+c(\beta+\mu)(\mu-\alpha )>0 \cdot
\end{equation}
For $(\mu-\alpha )>0$, both roots of $m_{0}$ will be negative. It means for any positive value of $m_{0}$, the required condition for the extinction of piracy will hold. For  $(\mu-\alpha )<0$, one of the roots will be positive and that will be the threshold value of $m_{0}$, denoted by $m_{0_{th}}$. If the value of intrinsic social awareness is more than this positive root of the quadratic equation, society will be free from piracy. In Fig. \ref{fig:bif steady}(b), we can observe that beyond this threshold value, a fraction of unaware becomes 1 indicating the absence of piracy.
\vspace{-3mm}
\section{Heterogeneous Analysis}
Though the homogeneous model predicts the steady state of the system, it does not give any information about the time evolution of a particular node. In this section, we will analyze the diffusion of piracy habits in heterogeneous populations incorporating its network structure in the differential equation model.  
\vspace{-2mm}
\subsection{Degree Block Approximation }
In terms of density functions, equation \ref{diff_eq:uba_with_media_model_3}  can be written as
\begin{eqnarray}
u_k'&=&\mu -\alpha _{n} k   (1-\Theta) u_{k} \psi_{b}-\mu u_{k},  \nonumber \\
b_k'&=&\alpha _{n} k  (1-\Theta) u_{k} \psi_{b}-\rho_{n} k b_{k} \psi_{a}+\beta a_{k}-\gamma m b_{k}-\mu b_{k},  \nonumber \\
a_k'&=&\rho_{n} k b_{k} \psi_{a}-\beta a_{k} +\gamma m b_{k} -\mu a_{k}, \label{diff_eq_uk_bk_ak_model_3_with_media}\\
m'&=&\phi b- \phi_{0}(m-m_{0})\cdot \nonumber
\end{eqnarray}
Diffusion in networks will depend on the network's degree distribution, as opposed to the homogeneous method. Hence, In a heterogeneous setting, here $u$, $b$, and $a$ have been modified to  $u_k$, $b_k$, and $a_k$ as the fraction of unaware, bootlegger, aware nodes having degree k. And also here we used the rate parameters as $\rho_n$ and $\alpha_n$ instead of $\rho$ and $\alpha$ of heterogeneous analysis. The value of $\rho_n$ should be the average piracy habit spread in the heterogeneous approach is equal to the overall spread by a bootlegger in the homogeneous approach.  But, the impact of mass media has been considered the same for all nodes irrespective of their degrees. That's why in the expression of $m'$, we have used $b$, not $b_k$. Here, $b$ signifies the fraction of bootleggers in the entire population.  Multiplying the first three equations of the coupled eq. \ref{diff_eq_uk_bk_ak_model_3_with_media} by $\frac{kp_{k}}{\langle k \rangle}$ and summing over $k$, we get
\begin{eqnarray}
\psi_{u}'
&=&\mu -\alpha _{n} \sum_{k} \frac{k^{2} p_{k}}{\langle k \rangle }u_{k} \frac{c}{c+m}\psi_{b}-\mu \psi_{u}, \nonumber \\
\psi_{b}'
&=&\alpha _{n} \sum_{k} \frac{k^{2} p_{k}}{\langle k \rangle }u_{k} \frac{c}{c+m} \psi_{b}-\rho_{n} \sum_{k} \frac{k^{2} p_{k}}{\langle k \rangle }b_{k} \psi_{a}+\beta \psi_{a}\nonumber \\ &\phantom{}&-\gamma m \psi_{b} -\mu \psi_{b} , \label{diff_eq_theta_uba_model_3_with_media}\\
 \psi_{a}'
&=&\rho_{n} \sum_{k} \frac{k^{2} p_{k}}{\langle k \rangle }b_{k} \psi_{a}-\beta \psi_{a}+ \gamma m \psi_{b} -\mu \psi_{a} \cdot \nonumber
\end{eqnarray}
Here $p_k$ stands to be the fraction of nodes in the network with degree $k$ and $\langle k \rangle$ is the average degree.
\subsection{Early Stage Analysis}
\label{early_stage_3A}
Analyzing the dynamics from the early stage where very few people are aware of using pirated versions we are approximating $u_{k}$ by 1 and $b_{k}$ as well as $a_{k}$ is considered to be negligible. Using this approximation to linearize the nonlinear terms,  eq. \ref{diff_eq_theta_uba_model_3_with_media} is simplified to
\begin{eqnarray}
\psi_{u}'
&=&\mu -\alpha _{n}  \frac{\langle k^{2} \rangle}{\langle k \rangle } \frac{c}{c+m}\psi_{b}-\mu \psi_{u}, \nonumber \\
\psi_{b}'
&=&(\alpha _{n}  \frac{\langle k^{2} \rangle}{\langle k \rangle }\frac{c}{c+m}-(\gamma m +\mu))\psi_{b}+\beta \psi_{a},  \label{diff_eq_theta_uba_early_stage_model_3}\\
\psi_{a}'
&=& -(\beta+\mu) \psi_{a}+ \gamma m \psi_{b} \cdot \nonumber
\end{eqnarray}
Last two equations of the coupled eq. \ref{diff_eq_theta_uba_early_stage_model_3} forms a system of simultaneous linear differential equations with constant coefficients.
\begin{eqnarray}
\psi_{b}'
&=& C_{1}\psi_{b}+C_{2} \psi_{a},\;\;\;\;
\psi_{a}'
= C_{3} \psi_{a}+ C_{4} \psi_{b} \cdot
\label{diff_eq_theta_a_simultaneous_equation}
\end{eqnarray}
We observe that the necessary condition for initial growth in class $B$ is  $C_{1}C_{3}<C_{2}C_{4}$ \cite{sayantari2020ensuring}. Substituting the expression of all these constant terms, the condition modifies to
\begin{equation}
\frac{\alpha _{n}}{\mu} (\frac{c}{c+m})(\frac{\beta+\mu}{\beta+\mu+\gamma m})>\frac{\langle k \rangle}{\langle k^{2} \rangle} \cdot \nonumber
%\label{R_m_condition_with_media}
\end{equation}
Replacing $\alpha _n$ by $\frac{\alpha }{\langle k \rangle}$, we get
\begin{equation}
\frac{\alpha }{\mu} (\frac{c}{c+m})(\frac{\beta+\mu}{\beta+\mu+\gamma m})= \mathcal{R}_m>\frac{\langle k \rangle ^2}{\langle k^{2} \rangle} \cdot \nonumber
\label{R_m_condition_with_media}
\end{equation}
The right-hand side of the inequality is \textit{epidemic threshold} in terms of network parameters. The first and second moments of degree distribution will depend on the network structure.
%R_{m}<\frac{\langle k \rangle}{\langle k^{2} \rangle}
%\end{equation}
\begin{figure}
\begin{center}
\includegraphics[width=0.8\linewidth]{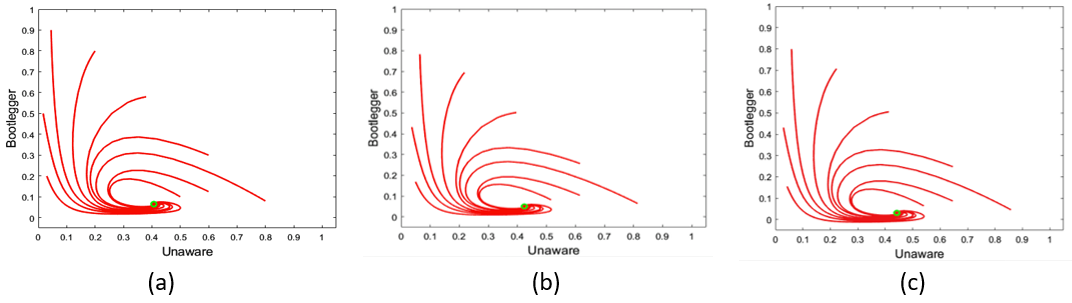}
 \caption{Temporal variation of $u$ and $b$ in case of an endemic steady state with different initial conditions the for the parameter set $\mu= 0.05$, $\alpha = 2$, $\beta= 0.01$, $\rho= 0.3$, $\gamma= 0.08$, $m_0=4$, $c=5$, $\phi=0.5$ and $\phi_0=0.1$ in case of (a) homogeneous setting; (b) Random network; (c) Jazz network. Value of $\mathcal{R}_m$ for the considered parameter set is 3.5, which is greater than 1.}
\label{flow_diagram}
\end{center}
\end{figure}
\subsection{Steady State Analysis }
In steady state, rate of change of $u_{k}$, $b_{k}$, $a_{k}$ and $m$ will be zero. Equating the last equation of the coupled eq.  \ref{diff_eq_uk_bk_ak_model_3_with_media} to zero we get $m=m_{0}+zb$ where $z=\frac{\phi}{\phi_{0}}$. Using this expression of $m$ in eq.  \ref{diff_eq_uk_bk_ak_model_3_with_media}, we get
\begin{equation}
u_{k}={\scriptstyle{\frac{\mu (c+m_{0}+bz)}{\mu (c+m_{0}+bz)+\alpha _{n} k c \psi_{b}}}} ; \
a_{k}={\scriptstyle{\frac{\rho_{n} k \psi_{a}+ \gamma(m_{0}+bz)}{\beta+\mu}}}b_{k}\cdot \nonumber
\label{with_media_uk_and_ak_in_theta_a_b}
\end{equation}
Substituting these expressions of $u_{k}$ and $a_{k}$ in $( u_{k}+b_{k}+a_{k}=1)$, we get
\begin{equation}
b_{k} ={\scriptstyle{(\frac{\beta+\mu}{\beta+\mu+\rho_{n} k \psi_{a}+\gamma(m_{0}+bz)})(\frac{\alpha _{n} k c \psi_{b}}{\alpha _{n} k c \psi_{b}+\mu (c+m_{0}+bz)})}}\cdot \nonumber
\label{with_media_bk_in_theta_a_b}
\end{equation}
Multiplying above equation by $\frac{kp_{k}}{\langle k \rangle}$ and
summing over $k$ we get
\begin{equation} 
 \psi_{b} = \sum_{k} \frac{ k p_{k}(\beta+\mu)\alpha _{n} k c  \psi_{b} }{\parbox{2.3in}{$\{\langle k \rangle (\beta+\mu+\rho_{n} k \psi_{a}+ \gamma (m_{0}+bz))$ %\\\hspace*{1.6cm}
 $( \alpha _{n} k c\psi_{b}+\mu(c+m_{0}+bz))\}$}}\cdot 
\label{with_media_theta_b_self_consistency}   
\end{equation}
This is a self-consistency equation of the form $\psi_{b}=f(\psi_{b}, \psi_{a})$ having 0 as an obvious solution. At $\psi_{b}=1$, which also implies $b=1$
 \begin{equation}
f(1, \psi_{a}) = \sum_{k} \frac{k p_{k}(\beta+\mu)(\alpha _{n} k c)}{\parbox{2.3in}{$\{\langle k \rangle (\beta+\mu+\rho_{n} k \psi_{a}+\gamma (m_{0}+z))$ \\\hspace*{1.6cm}$( \alpha _{n} k c + \mu (c+m_{0}+z))\}$}}\cdot
\label{with_media_f_1_Theta_a}
\end{equation}
Observing the numerator and denominator of the eq. \ref{with_media_f_1_Theta_a}, we can see that for every term in the numerator, there is a corresponding larger term in the denominator. Hence, $f(1, \psi_{a})$ will surely be less than 1. To have a solution of eq. \ref{with_media_theta_b_self_consistency} in the interval $\psi_{b}=(0,1)$, slope of the function $f$ must be greater than 1 at the point $(\psi_{b}=0, \psi_{a}=0)$. The slope of the function is
\begin{eqnarray}
\frac{\partial f(\psi_{a},\psi_{b})}{\partial \psi_{b}}&=& \frac{(\beta+\mu)\alpha _{n} c}{\langle k \rangle (\beta+\mu+\rho_{n} k \psi_{a}+\gamma (m_{0}+bz))}\times \nonumber \\ &\phantom{}&\sum_{k} \frac
{\mu (c+m_{0}+bz) k^{2} p_{k}}{(\mu(c+m_{0}+bz)+\alpha _{n} k c\psi_{b})^{2}}\cdot \nonumber
%\label{with_media_slope_f_Theta_a_b}
\end{eqnarray}
At point $(\psi_{b}=0, \psi_{a}=0)$, value of the slope is
 \begin{equation}
\frac{(\beta+\mu)\alpha _{n} c \langle k^{2} \rangle }{(\beta+\mu+\gamma m_{0})\mu (c+m_{0}) \langle k \rangle}\cdot  \nonumber
\end{equation}
 Hence, the required condition to have a desired solution for the eq. \ref{with_media_theta_b_self_consistency} is
\begin{equation}
\frac{\alpha _{n} (\beta+\mu)c }{\mu (\beta+\mu+\gamma m_{0})(c+m_{0}) }>\frac{\langle k \rangle}{\langle k^{2} \rangle} \cdot \nonumber
\end{equation}
Substituting $\alpha _n$ by $\frac{\alpha }{\langle k \rangle}$, the condition is equivalent to
$
R_{m}>\frac{\langle k \rangle ^2}{\langle k^{2} \rangle} \nonumber
$
which is the same as obtained in Sec. \ref{early_stage_3A}.
\section{Numerical Results}
In this section, we are going to discuss the results of the spread of piracy habits after applying mass media awareness in society. We will compare the results of the homogeneous and heterogeneous analysis. In the heterogeneous part, along with model networks, results over real networks will also be discussed.

\subsection{Simulation of Homogeneous Model}
{In Sec. \ref{stability_3A}, we found that after mass media awareness, people with a habit of piracy will exist only beyond $\mathcal{R}_m=1$. If we consider the reproduction number without media campaigns, $\mathcal{R}_m$, includes two multiplicative factors: $\frac{\beta+\mu}{\beta+\mu+\gamma m_0}$ and $\frac{c}{c+m_{0}}$, which are always less than $1$. Hence, effective implementation of mass media awareness programs makes it difficult to cross the epidemic threshold and enter the endemic region.
\newline
Temporal evolution in $u-b$ plane for different initial points have been shown in Fig. \ref{flow_diagram}(a), for $\mathcal{R}_m>1$. We can observe that at a steady state, the fraction of bootleggers in the population is very less.

\begin{table}
\centering
\caption{Steady state values of different classes for different networks in the presence of mass media awareness.\\} 

\label{real_net_compare_3A}
{\small{
\begin{tabular}{|c|l|l|l|l|l|l|}
\hline
\begin{tabular}[c]{@{}c@{}}Steady state\\ fraction\end{tabular} & \multicolumn{1}{c|}{\begin{tabular}[c]{@{}c@{}}Homogeneous\\
Setting\end{tabular}} & \multicolumn{1}{c|}{\begin{tabular}[c]{@{}c@{}}Random\\ Network\end{tabular}} & \multicolumn{1}{c|}{\begin{tabular}[c]{@{}c@{}}Jazz\\ Network\end{tabular}} \\ \hline
$u^{\star}$
& 0.407
& 0.428                                                                          & 0.452                                                                                                                                                                                                                                                                                                     \\ \hline
$b^{\star}$
& 0.061
& 0.060                                                                         & 0.058                                                                                                                                                    \\ \hline
$a^{\star}$                                                               & 0.507                                                                         & 0.483                                                                             & 0.402                                                                                          \\ \hline
\end{tabular}
}}
\end{table}
\subsection{Simulation over Model Networks}
 Similar to homogeneous cases, we can notice the significant decrement in the steady state fraction of bootleggers in the case of model networks also. Temporal evolution for the random network has been shown in Fig. \ref{flow_diagram} (b). The steady-state value in the case of a random network almost matches the homogeneous scenario. Error in the steady-state fractions of different classes is bounded by 2 \% for the considered parameter set in the case of a random network. Qualitatively similar results could be expected For a scale-free network as well. However, the variety in its structure may cause the error to be relatively substantial. \\
\begin{figure}
    \centering
    \includegraphics[width=\linewidth]{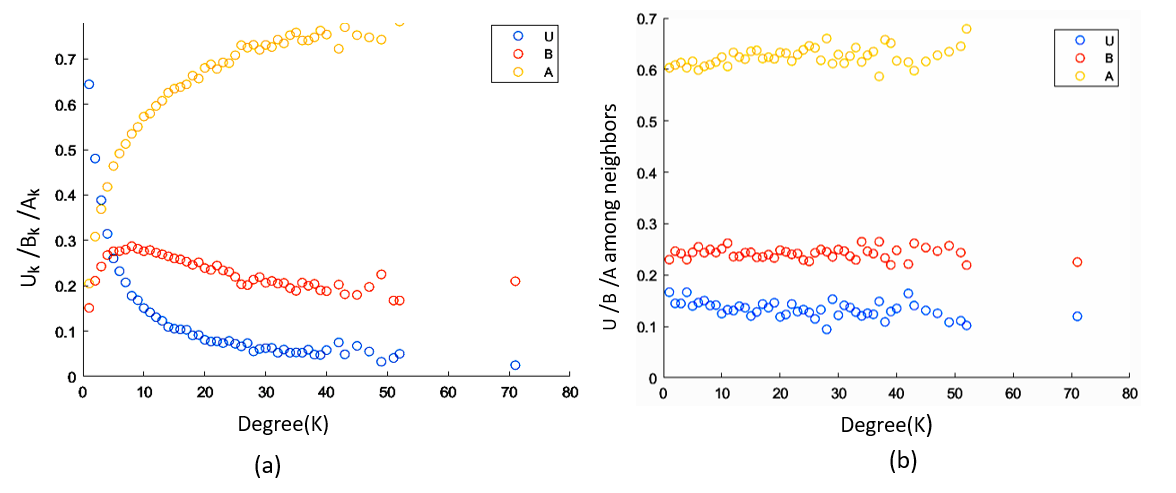}
    \caption{(a) Degree wise fraction of u, b and a concerning degree k in case of a random network at steady-state (b) Fraction of u, b and an in the neighborhood of a node with degree k in case of a random network}
    \label{degree based}
\end{figure}

There will be more believers nearby a node with a lot of neighbors. When more people are attempting to disseminate the story to a person, there are obviously greater chances for that person to hear it.

To understand the impact of the diffusion process from the perspective of nodes with different degrees in the presence of media in a random network, we have plotted the degree-wise steady-state fraction of different classes in Fig. \ref{degree based} (a). It can be observed from the figure that the fraction of bootleggers is monotonically increasing with degree. Meanwhile, the probability of a node being in an unaware class is decreasing with the nodal degree. To understand the reason behind this observation, we have plotted the fraction of different classes in the neighborhood of a node in Fig. \ref{degree based}(b). It can be seen that this fraction is about the same for all degrees in the random network. This indicates that, statistically, the proportion of neighbors who may disseminate piracy habits to a node is constant across all nodes. But that does not imply that there are the same amount of bootleggers everywhere. It indicates that higher-degree nodes are surrounded by more bootleggers. That is why they are more prone to piracy habit. A similar pattern can be expected in the case of scale-free networks as well. Hence the impact of the diffusion process is more on higher degree nodes. These nodes have a  high probability to undergo a transition from one class to another and make other nodes to getting connected to them. Hence, they are crucial in the diffusion process.

\subsection{Simulation over Real Networks}

Here, we've applied the internet piracy contagion concept to a real-world jazz network, or the network of collaboration amongst jazz musicians. Here, each musician is a node, and the connection between them denotes that they played in a band together. The information was gathered from the KONECT (The Koblenz Network Collection) database to comprehend its impact on actual social interaction settings. Similar to the homogeneous part and random network the temporal evolution in the $u-b$ plane, for the jazz network for this endemic scenario, has been shown in Fig. \ref{flow_diagram}(c). \\
All the results shown for the steady-state values of different classes under the same parameter set, for homogeneous, random, as well as for the Jazz real network have been listed in Table \ref{real_net_compare_3A}.\\ Here, it is important to observe that the steady-state values in homogeneous and heterogeneous populations qualitatively resemble each other quite a bit. Even though the numerical numbers in real networks vary depending on the topology and network features, the physical interpretations and conclusions made based on a homogeneous situation are nonetheless unquestionably true. Furthermore, in both situations, the impacts of the rate parameters on diffusion are equal.

\begin{figure}
\centering
    \includegraphics[width=0.9\linewidth]{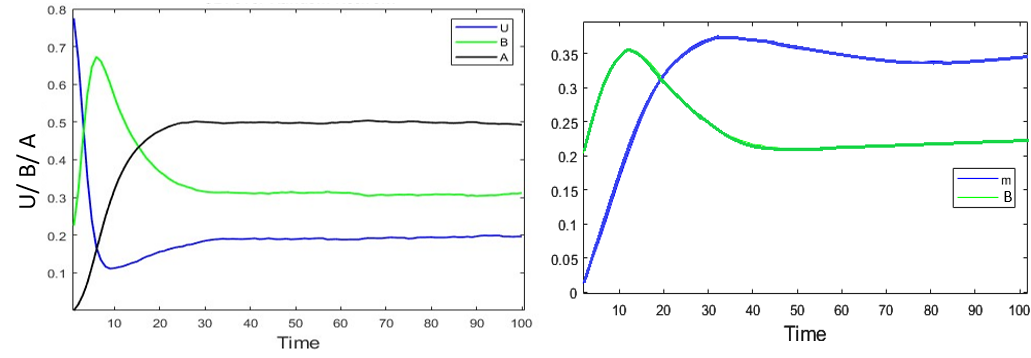}

\caption{(a)Temporal evolution of $u$, $b$, and $a$ in the presence of mass media awareness with an initial condition  (0.75,0.22,0) (b) Similar Variation of media level and population of bootleggers with time. Parameter set for both the plots are $\mu=0.05$, $\alpha =2$, $\beta=0.2$, $\rho=0.3$, $\gamma=0.08$, $m_0=4$, $c=5$, $\phi=0.5$, and $\phi_{0}=0.1$.}
\label{uba_vs_time}

\end{figure}

We have also shown the time evolution of $u$, $b$, and $a$ in the case of a random network in Fig \ref{uba_vs_time} (a) for a particular initial condition and seen that after some time all the population is becoming steady. In Fig. \ref{uba_vs_time}(b), we have shown the variation of media awareness level and the variation of bootleggers population in a single plot to emphasize that for the success of a media awareness campaign, the level of media must be adjusted proportionally with population of bootleggers in the society. And, here we observed that for some point particularly where the media implication has started from exactly that point the bootlegger population started decreasing. So we can conclude that the implication of a media awareness program could be an effective way to control this online piracy contagion.  

\section{Conclusion \& Discussion} 
Though several attempts are made to analyze outbreaks using homogeneous and heterogeneous approaches independently, it is not well-explored what are the similarities and dissimilarities between these two approaches. Upon first inspection, both these approaches seem quite different. Homogeneous modeling is based on differential equations and relies on strict assumptions about homogeneity, whereas heterogeneous analysis
is based on simulations carried over a heterogeneous network structure. Conversely, homogeneous analysis is expected to run fast whereas heterogeneous analysis may take hours to give results, depending on the size of the network and complexity of the interactions. In this paper, we aim to model the social phenomena of online piracy, which exhibits epidemic-like diffusion due to its strong reliance on peer influence. Nowadays, online piracy i.e., the illegal act of copying, duplicating, and sharing digital work without the permission of copyright holders is becoming a burning issue, globally, for the digital industry. In our study, the online piracy habit has been modeled and analyzed from the perspective of the effectiveness of mass media campaigns to develop awareness among individuals. We used ODE models as well as graph-theoretical treatment with differential equations on the network to figure out ways to control the habit of piracy in society. The presence of media campaigns helps to maintain the number of aware people in the population and directly restricts the spread of this piracy habit. To examine the similarities and differences for both the mean-field and network approaches, related to this pirate dynamic, we used both the homogeneous and heterogeneous methodologies. Here, we discovered the parameter threshold that regulates the presence of piracy states and detected the bifurcations in the system. To demonstrate the presence of online piracy prevalence and control, numerical simulations for this dynamic were carried out on random as well as actual online social networks. Apart from detailed mathematical and computational analyses, similarities and dissimilarities between the two approaches have also been explicitly quantified. We observe that while there is
certain physical information that can be achieved from both approaches, each of them explicitly reveals specific characteristics about the flow in society which can be crucial to develop promotional or inhibiting activities.
%ODE-based analysis followed by bifurcation analysis where we observed the steady state               

\bibliographystyle{elsarticle-num} 
\bibliography{cas-refs}

\end{document}